\documentclass[pre,floatfix,twocolumn,showpacs,a4paper,nofootinbib]{revtex4}
\usepackage{amsmath}
\usepackage{amsfonts}
\usepackage{mathrsfs}
\usepackage{epsfig}
\usepackage{graphicx}
\usepackage{dcolumn}
\usepackage{amsmath}

\def\be{\begin{equation}}
\def\ee{\end{equation}}

\begin{document}

\title[The statistical laws of popularity: Box-office dynamics of
movies]{The statistical laws of popularity: Universal properties of
the box office dynamics of motion pictures}

\author{Raj Kumar Pan$^{1,2}$ and Sitabhra
Sinha$^{1}$}

\address{$^1$The Institute of Mathematical Sciences, CIT Campus, Taramani, Chennai 600113, India\\
$^2$Department of Biomedical Engineering and Computational Science,
School of Science and Technology, Aalto University,
P.O. Box 12200, FI-00076 AALTO,
Finland}
\begin{abstract}
Are there general principles governing the process by which certain
products or ideas become popular relative to other (often
qualitatively similar) competitors~? To investigate this question in
detail, we have focused on the popularity of movies as measured by
their box-office income. We observe that the log-normal distribution
describes well the tail (corresponding to the most successful movies) 
of the empirical 
distributions for the total income, the income on the opening week, as
well as,
the weekly income per
theater. This observation suggests that popularity may
be the outcome of a linear multiplicative stochastic process.
In addition, the distributions of the total income and the opening income
show a bimodal form, with the majority of movies either performing
very well or very poorly in theaters.
We also observe that the gross income per theater for a movie 
at any point during its lifetime is, on
average, inversely proportional to
the period that has elapsed after its release. 
We argue that (i) the log-normal nature of the tail, (ii) the bimodal
form of the overall gross income distribution, and
(iii) the decay of gross income per theater with time as a power law,
constitute the fundamental set of {\em stylized facts} (i.e.,
empirical ``laws") 
that can
be used to explain other observations about movie popularity. We show
that, in
conjunction with an assumption of a fixed lower cut-off for income per theater
below which a movie is withdrawn from a cinema, these laws can be used
to derive a Weibull distribution for
the survival probability of movies which agrees with empirical data.
The connection to extreme-value distributions
suggests that popularity can be viewed as a process where a product
becomes popular by avoiding ``failure" (i.e., being pulled out
from circulation) for many successive time periods.
We suggest that these results may apply beyond the particular
case of movies to popularity in general.
\end{abstract}

\pacs{89.65.-s, 05.65.+b, 02.50.-r, 89.75.Da}
\maketitle

\begin{quote}
{\small ``Now that the human mind has grasped celestial and terrestrial physics, -
mechanical and chemical, organic physics, both vegetable and animal, - there
remains one science, to fill up the series of sciences of observation, - 
Social physics. This is what men have now most need of \ldots" - 
Auguste Comte, {\em The Positive Philosophy of Auguste Comte}, tr. H Martineau,
New York: Calvin Blanchard, 1855, p. 30.}
\end{quote}

\section{Introduction}
As indicated by the above quotation from August Comte, the 19th century French philosopher
and founder of the discipline of sociology, the idea of using the principles
of physics to analyze and understand social phenomena is not new. Indeed, as
Philip Ball has recently pointed out~\cite{Ball03}, one can trace
connections between physics
and the study of socio-political phenomena back to the 16th century with
the English philosopher, Thomas Hobbes. 
Hobbes, who had contributed, among other topics, 
to the physics of gases as well as to the theory of political science, believed that
the deterministic principles of Galilean physics could be used to frame the laws governing
not only the behavior of particles of matter but also that of particles of society, i.e., 
human beings. At about the same time an associate of Hobbes, William Petty, had appealed for an
empirical study of society in the manner of physics, by measurement
of quantifiable demographic and economic variables. This anticipated the use
of statistical techniques to study society, which was eventually developed
by the 19th century Belgian mathematician, Adolphe Quetelet.
Quetelet's book {\em Essai de physique sociale} (1835) outlined the concept
of identifying the statistical characteristics of the ``average man" by collecting
data on a large number of measurable variables~\cite{Quetelet42}. 
In fact, it is here 
the term ``social physics" was used possibly for the first time to
indicate the quantitative study
of empirical data collected from a population. Its goal was to elucidate the
statistical laws underlying social phenomena such as the incidence of crime or the rates
of marriage. Quetelet used the 
Gaussian distribution, which had previously been introduced in the 
context of analysis of astronomical observations,
to characterize many aspects of human behavior. Not surprisingly, this
aroused great controversy among his contemporaries. There was an
intense debate as to whether this meant that the freedom of choice for an 
individual was illusory. However, Quetelet was quite clear that this only implied
an overall average tendency towards a certain behavior, rather than
predicting
how a specific individual will behave under a given set of circumstances. It is instructive
that even today much of the opposition to the application of physics to study society
is based on an erroneous impression that such an approach implies exact
predictability of individual behavior.

Given the long historical connection between physics and the study of society,
it is therefore surprising why more than a century has passed after the 
development of statistical physics 
before it was
applied in a serious and sustained manner to
analyze various social, economic and political phenomena.
There have, of course, 
been several individual attempts in this direction earlier (e.g., see
Ref.~\cite{Mantegna05}).
The appealing parallels between the collective behavior
of a large assembly of individuals, each of whose actions in isolation are almost
impossible to predict, and that of a large volume of gas, where the dynamics of 
individual molecules are difficult to specify, have been noted even by
non-physicists. For example, the science fiction writer, Isaac Asimov, has used
this analogy as the basis for the fictional discipline of ``psycho-history" in his 
{\em Foundation} series of novels where 
statistical laws describing the behavior of entire populations are used to
predict the general historical outcome of events. It is of interest to note in
this context that the framework of kinetic theory of gases has been used
to explain certain universal socio-economic distributions, most
notably,
the distribution of affluence, measured in terms of either individual income or wealth~\cite{Chatt07,Sinha09}. 

However, it is only recently that there has been
a large and concerted initiative by statistical physicists to explain social phenomena 
using the tools of their trade~\cite{Castellano09}. This may be partly due to
lack of sufficient quantities of high-quality data related to society. 
Indeed, the advances made in {\em econophysics}, i.e., the study of economic
phenomena using the principles of physics, especially of properties 
pertaining to
financial markets, have been attributed to the availability of large volumes
of data. This has made it possible to quantitatively
substantiate the existence of universal properties in the behavior of
such systems, e.g., the {\em inverse cubic law} of the distribution of price 
fluctuations~\cite{Gopikrishnan98,Lux96,Pan07}. These statistical laws, often referred
to as ``stylized facts" in the economic literature, 
are seen to be invariant with respect to different realizations
of the system being studied.

More importantly, the existence of statistically universal behavior
in socio-economic phenomena suggests that it may indeed be possible to
explain them using the approach of physics, where the general
properties at the systems-level
need not depend sensitively on the microscopic details of its
constituent elements. Indeed, it was the empirical observation that critical exponents
of different physical systems are independent of the specific physical properties
of their elementary components that gave rise to the modern theory of critical
phenomena in statistical physics. The expectation that a similar path
will be followed by the physics of socio-economic phenomena has driven
the quest for empirical statistical laws governing such behavior, 
over the past decade and half~\cite{Mantegna00}.

One of the areas of social studies where the search for statistical universal
properties can be most fruitful is the emergence of collective decisions
from the individual choice behavior of the agents comprising a group. Many such decisions
have a binary nature, e.g., whether to cooperate or defect, to drop out of school, 
to have children out-of-wedlock, to use drugs, etc., which are reminiscent of the 
framework of spin models used to study cooperative phenomena in statistical 
physics~\cite{Durlauf99}.
The traditional
economic approach to understand how such choices are made has been that 
every individual arrives
at a decision based on the maximization 
of an utility function specific to him/herself and relatively independent of 
how other agents are behaving. 

However, it has been clear for quite some time, e.g., through the work of Schelling
on housing segregation~\cite{Schelling78}, that in many, if not most such
choice processes, an individual's decision is influenced by those of its peers or 
members of its social network~\cite{Wasserman94,Vega-Redondo07}. This interactions-based
approach, where the social context is an important
determinant of one's choice, provides an alternative to utility maximization by individual
rational agents in understanding several social phenomena. In
particular, such an approach may be necessary for explaining the sudden
emergence of a widely popular product or idea which is otherwise
indistinguishable
from its competitors in terms of any of its observable qualities. While traditional
economic theory would hold that this suggests that there is a term in the utility function
which corresponds to an {\em unobservable} property differentiating the popular entity 
from its competitors, 
there is no way of testing its scientific validity 
as such an hypothesis cannot be subjected to empirical verification. 
The alternative interactions-based
viewpoint would be that, despite the absence of any intrinsic advantage initially, the 
chance accumulation of a relatively larger number of adherents early on would produce 
a slight relative bias in favor of adoption of a specific idea or product. Eventually,
through a process of self-reinforcing or positive feedback via interactions, an inexorable  
momentum is created in favor of the entity that builds into an enormous advantage in
comparison to its competitors (see, e.g., Ref.~\cite{Arthur90}).

While most such decisions do not have life-altering repercussions, the
empirical study of collective choice dynamics can nevertheless result
in formulation of statistical laws that may provide us with better
understanding of, for instance, how financial manias sweep through
apparently rational and highly intelligent traders or what leads to
publicly sanctioned genocides in civilized societies.  A relatively
innocuous example of such choice behavior is seen in the emergence of
movies that become extremely popular, often colloquially referred to
as ``superhits", and its
flip-side, the ignoble departure of intensely promoted movies which
nevertheless fail miserably at the box-office. Frequently, it is not
easy to see what quality differences (if any) are responsible for the
runaway success of one and the mediocre performance of the 
others\footnote{The relative absence of correlation between quality and
popularity has also been observed experimentally in an artificial
``music market" where participants downloaded previously unknown songs
either with or without knowledge of the choices made by other
participants~\cite{Salganik06}.}. It
is thus a fecund area to search for signatures of statistical   laws
of popularity that can give us some indications of the essential
dynamics underlying the emergence of collective decisions from
individual choice.

In this paper, we have presented our results of a detailed analysis of the box-office
performance of motion pictures released in theaters across the USA over the past several years, 
which are corroborated by data from India and Japan. 
The key finding reported here is that such choice dynamics has three prominent
universal features which are invariant with respect to different periods of observations
(and hence, different ensemble of movies). First, the gross income of
a movie in its opening week, normalized by the number of theaters in which 
it is being shown, follows a log-normal distribution. Second, the
number of theaters a movie is released in has a bimodal distribution.
Together, these two properties account for the bimodal log-normal
nature of the opening gross distribution for all movies released
in a particular year. Further, we note that the total gross income of a movie
over its entire theatrical run also follows a distribution that can be
understood as a superposition of two log-normal distributions.
The similarity between the opening and total gross distributions is related to
the third universal feature of movie popularity, viz., the
average gross weekly income per theater of a movie decreases 
as an inverse linear function of the time that has
elapsed after its release. Together these three properties 
determine all observable characteristics of the movie ensemble, including
the distribution of their lifetimes. 

In section 2, we introduce
the terminology used in the rest of the paper, discussing specific 
features of movie popularity and its quantifiable measures. This is
followed
by a short description of the data-sets that we have used for our
analysis in section 3.
In section 4, which describes our main results, first the properties of the
distributions of the gross income, both for the opening week, as well as, over 
the total duration that a movie is shown in theaters, is discussed. 
We follow this up with results 
on the distribution of the number of theaters in which a movie is released, and try to
see if any correlation exists between the opening and the overall
performance of the motion picture at the box office. The
time-evolution of a movie's
box-office performance is analyzed next.
Our results
show scaling in the decay of the income of a movie per theater as a function
of time. In section 5, we discuss the distribution of
persistence time (i.e., the duration up to which a movie is shown at theaters) 
which exhibits the properties
of an extreme value distribution. This observation leads us to suggest that
popularity can be viewed as a sequential survival process over many
successive time periods, 
where a successful product
or idea is one that survives being withdrawn from circulation for far
longer than its competitors.
We conclude with a conjecture that the
invariant properties observed for movie popularity 
may also be relevant for other instances of popularity.

\section{Measuring movie popularity}
For movies, as for most other products competing for attention by potential
consumers/adopters, it is an empirically observed fact that only a very few 
end up dominating the market. As Watts points out ``\ldots
for every {\em Harry Potter} and {\em Blair Witch Project} that explodes out of 
nowhere to capture the public's attention, there are thousands of books,
movies, authors and actors who live their entire inconspicuous lives beneath
the featureless sea of noise that is modern popular
culture"~\cite{Watts03}.  In fact, it is an oft-quoted fact about the
motion picture industry that the majority of movies released every
year do not attract enough viewers. The following comment made about
Bollywood, the principal Indian film production and distribution
system, applies also to motion picture industries worldwide: 
``fewer than 8 out of the more than 800 films made each year [in
Bollywood] will make serious money"~\cite{Torgovnik03}.

It may be worth mentioning here that popularity can arise through different
processes. For example, a product can become a runaway success immediately
upon release, its popular appeal presumably being driven by a saturation advertising
campaign across different media that precedes the release. This has been the
dominant marketing strategy for successful big budget movies 
which are collectively referred to as ``blockbusters". 
In contrast, products that are unsuccessful, referred to as
``bombs" or ``flops" in the context of movies, get a relatively poor 
reception on being released in the market. For such movies, a bad opening
week is usually followed by ever-declining ticket sales resulting in a quick
demise at the box-office. However, an alternative scenario is also
possible where, after a modest opening week performance, a movie
actually shows an increase in its popularity over subsequent periods.
This process is thought to be driven by word-of-mouth promotion of
the product through the social network of consumers who influence the choice of their
friends and acquaintances. As more people adopt or consume the product, its popular appeal
is increased further in a self-reinforcing process. These
kinds of popular products, often termed ``sleeper hits" in the context
of movies, are much less frequent relative to the blockbusters and the
bombs.

As in any quantitative study of the emergence of collective decision
concerning
the adoption of a product or an idea, the 
first question we need to resolve is how to measure popularity. 
While in some cases it may be rather obvious, as for example, the number
of people driving a particular brand of car or practising a particular religion,
in other cases it may be difficult to identify a unique measurable property
that will capture all aspects of popularity. In particular, the popularity
of movies can be measured in a number of ways. For example, we can look
at the average ratings given by critics in reviews published in various media,
web-based voting in different movie-related online forums, the cumulative 
number of DVDs rented (or bought) or the income from the initial theatrical run of a 
movie in its domestic market.

Let us consider in particular the case of movie popularity as judged by
votes given by registered users of one of the largest online movie-related 
forums, the Internet Movie Database (IMDb)~\cite{imdb}. Voters can rate a
movie with a number between 1-10, with 1 corresponding to ``awful'' and 
``10'' to excellent. The cumulative distribution of the total votes
for a movie approximately fits a log-normal distribution towards the tail,
with the maximum likelihood estimates of the distribution parameters
being $\mu=8.60$ and $\sigma=1.09$~\cite{Pan06}.
However, the use of such scores as an accurate measure of movie popularity has
obvious limitations. For instance, the different scores may be just a result
of voters having differing amounts of information about the movies, with older,
so-called classic movies clearly distinguished from newly released
movies in terms of the voters' knowledge about them.
More importantly, as it does not cost anything to vote for a particular movie
in such online forums, the vital element of competition for viewers
that governs which product/idea will eventually become popular is
missing from this measure. Therefore, we will focus on the box-office
gross earnings of movies after they are newly released in theaters as
a measure of their relative popularity. As we are considering only
movies in their initial theatrical runs, the potential viewers have
roughly similar kind of information available about the competing
items. Moreover, such ``voting with one's wallet'' is possibly a more
honest indicator of individual preference for movies. The availability
of large, publicly accessible data-sets of the daily or weekly
earnings of movies in different countries in several movie-related
websites
has now made this kind of analysis a practical exercise.

\section{Data Description}
For our study we have concentrated on data from the three most prolific feature
film producing nations in the world: India, USA and Japan~\cite{Hesmondhalgh07}. 
For example, of the 4,603 feature films produced around the world in 2005, India produced
1,041, USA 699 and Japan 356\cite{screendigest06}.
While the US film industry, centered at Hollywood, has led the rest of the world for many
decades both in terms of financial investment and the revenues generated by the films
they make, India has been for a long time the largest producer of movies with a 
correspondingly high cinema admission. In fact, India
accounts for  almost a quarter of the total number of feature films made annually worldwide. 
Both India and USA have large domestic markets for their films, 
with locally produced movies having more than $90\%$ of the market share 
in each country.
In addition, the US film industry also dominates the international market, with
most cinemas around the world showing movies made in Hollywood.
  
For detailed information on the gross box office receipts for movies released across 
theaters in USA during the period 1999-2008, we have used the data available from the 
websites of {\em The Movie Times}~\cite{movietimes} and {\em The Numbers}~\cite{numbers}.
The opening and total gross incomes of 5,222 movies released over this
period have
been considered in our analysis, as well as the maximum number of theaters in which 
they were shown. This roughly corresponds to including the 500-600 top earning movies
for each year. We have used this data to obtain the overall distributions of
movie popularity measured in terms of income. To compare the performance of a 
movie with its production budget, the latter information was obtained for
1,420 movies released over the period 2000-2008. We have also looked at the
time-evolution of popularity, focusing on the approximately top 300 movies
each year for the period 2000-2004. This corresponds to a total of 1497 movies
over the five-year period, where the total gross receipt at the box
office for each movie being considered is
greater than $10^5$ USD. For these movies, we
collected information about the gross income on each
week during the time the film ran in theaters, as well as, the number of theaters
that a movie was shown on a particular week. 

For verifying the universal nature of the distributions obtained from the US data, we
also considered the gross income data of a total of 500 movies released across 
theaters in India during the period
1999-2008 (corresponding to the top 50 movies each year in terms of
their box-office income) 
that was obtained from the {\em IbosNetwork} web-site~\cite{ibos}.
We also collected income data for 1095 movies released in Japan over the period 2002-2008
from the web-site {\em BoxOfficeMojo}~\cite{boxofficemojo}, which roughly corresponds to
the 150 top earning films each year.

\section{Results}
\subsection{Distribution of Opening and Total Gross}
As mentioned earlier, we will consider the gross income of a movie after it is released
at theaters as a measure of its popularity, as this is directly
related to the number of people
who have been to a cinema to watch it. Thus, the total gross of a
movie ($G_T$), i.e., the
entire revenue earned from screenings over the entire period that it was shown in theaters,
is a reasonable indicator of the overall viewership. 
An important property to note about the distribution of popularity for
movies, measured
in terms of their income, is that it
deviates significantly from a Gaussian form in having a much more extended tail.
In other words, there are many more highly popular movies than we would expect from
a Normal distribution.
This immediately suggests that the process of emergence
of popularity may not be explained simply as an outcome of many agents 
independently making binary (viz., yes or no) decisions to adopt a particular choice,
such as, going to see a particular movie. As this process can be
mapped to a random walk, we expect it to result in a Gaussian
distribution which, however, is not observed empirically. 
To go beyond this simple conclusion and
identify the possible processes involved behind the emergence of
popularity,
one needs to ascertain accurately the true nature
of the distribution of gross income.

The total gross is, of course, an aggregate measure of the box-office performance
of a movie. Over the theatrical lifetime of a movie, i.e., the period between
the time a movie is released to the time that it is withdrawn from the
last theater it is being shown at, its viewership can show remarkable up- and 
down-swings. For example, two movies which have the same total gross can take
very different routes to this overall performance. One movie may have 
opened at a large number of theaters and generated a large gross in its 
opening week followed by rapidly declining box-office receipts 
before being withdrawn. The other movie could have been initially released 
at a smaller number of theaters but, 
as a result of increasing popularity among viewers, was then
subsequently released at more theaters and ran for a 
longer period, eventually garnering the same
total gross. Thus, considering the opening gross of a movie ($G_O$),
i.e., the box-office revenue on the opening week, can provide us with 
information
about its popularity which complements that we have from its total gross.
Indeed, the opening week is considered to be the most critical event
in the commercial life of a movie as is evident from the pre-release publicity 
efforts of major Hollywood studios to promote movies in an effort to guarantee large
initial viewership. This is because the opening gross is widely thought to signal the 
success of a particular movie. The observation that about 65-70$\%$ 
of all movies earn their maximum box-office revenue in the first week
of release~\cite{Vany99} appears to support this line of thinking. In the
following analysis, we will look at both the total and the opening gross. 
\begin{figure}
\begin{center}
\includegraphics[width=0.99\linewidth]{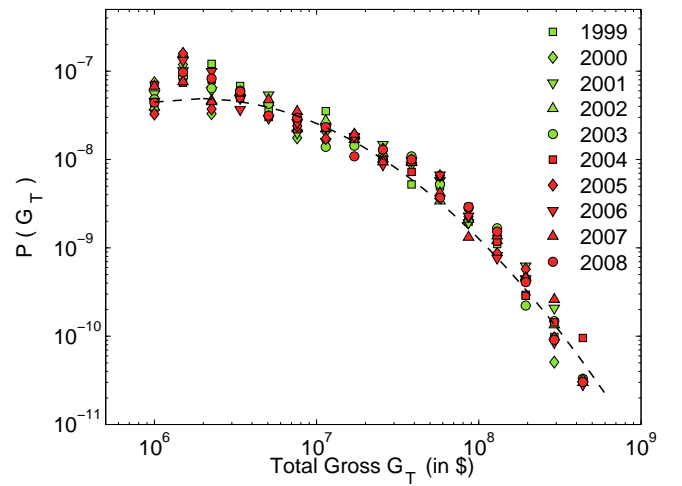}
\end{center}
\caption{Probability distribution of the total gross income, $G_T$,
for movies released across theaters in USA for each year during the period
1999-2008. The best fit for the aggregated data by a log-normal
distribution ({\em broken curve}) with maximum likelihood estimated parameters $\mu=16.607
\pm 0.062$ and $\sigma=1.471 \pm 0.042$, is shown for comparison. 
The estimates of the log-normal distribution parameters 
for each individual year lie within error bars of the aggregate distribution parameters.
}
\label{fig:tgross_us}
\end{figure}

\begin{figure}
\begin{center}
  \includegraphics[width=0.99\linewidth]{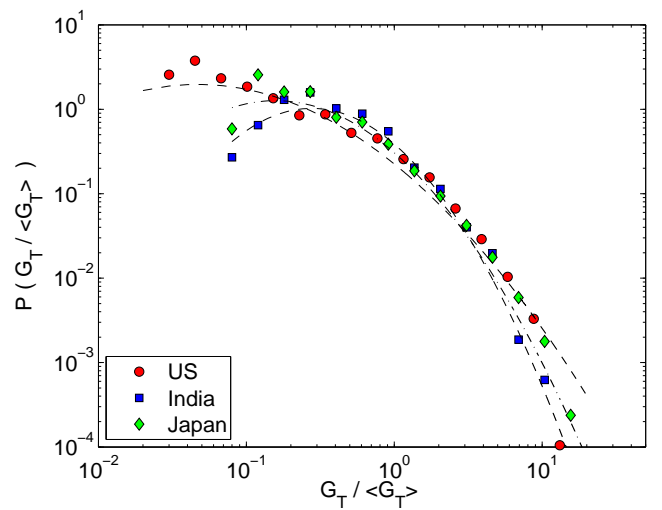}
\end{center}
\caption{Probability distribution of the total gross income, $G_T$,
scaled by the average total gross $\langle G_T \rangle$ of all movies
released across theaters in (a) USA during the period 1999-2008, (b)
India during the period 1999-2008 and (c) Japan during the period 2002-2008, 
each fit by a log-normal distribution ({\em broken curves}).
The maximum likelihood estimates of the parameters for the log-normal 
fits to the scaled total gross are 
(a) $\mu = -0.901 \pm 0.062$ and $\sigma = 1.471 \pm 0.042$ for
USA, (b) $\mu=-0.436 \pm 0.084$ and $\sigma=0.922 \pm 0.056$ for India
and (c) $\mu=-0.644 \pm 0.076$ and $\sigma=1.097 \pm 0.051$ for Japan.
For the unscaled data (i.e., $G_T$), the corresponding MLE parameters
are 
(i) $\mu=18.264 \pm 0.084$ and $\sigma=0.922 \pm 0.056$ for India, and
(ii) $\mu=15.725 \pm 0.076$   and $\sigma=1.097 \pm 0.051$ for Japan.
}
\label{fig:tgross_india}
\end{figure}
Previous studies of movie income 
distribution~\cite{Vany99,Sornette99,Vany03} had looked at limited data sets
and found some evidence for a power-law fit. A more rigorous analysis using data
for a much larger number of movies released across theaters in the USA
was done 
in Ref.~\cite{Sinha04}. 
While the tail of the distribution for the opening,
as well as, the total gross for movies, may appear to follow an approximate 
power law $P(I) \sim I^{-\alpha}$ with an exponent $\alpha \simeq
3$~\cite{Sinha04,Sinha05}, an even better fit is achieved with a
log-normal form (as shown in Fig.~\ref{fig:tgross_us}):
\begin{equation} 
P (x) = \frac{1}{x \sigma \sqrt{2 \pi}} e^{- ({\rm ln} x - \mu)^2/2
{\sigma}^2},
\label{eq:lognormal}
\end{equation}
where $\mu$ and $\sigma$ are parameters of the distribution, being the mean
and standard deviation of the variable's natural logarithm.
The maximum likelihood estimates of the log-normal
distribution parameters for the aggregated US gross income data are
$\mu=16.607$ and $\sigma=1.471$.  
The log-normal form has also been verified by us for the income distribution
of movies released in India and Japan (Fig.~\ref{fig:tgross_india}). 
It is of interest to note that a strikingly similar feature has been observed for 
the popularity of scientific papers, as measured by the number of their 
citations, where initially a power law was reported for the
probability distribution with exponent $\simeq 3$
but was later found to be better 
described by a log-normal form~\cite{Redner98,Redner04}. 
\begin{table*}[htbp]
\centering
\caption{Comparison between Kolmogorov-Smirnov statistics for the
fitting of the tails of total gross ($G_T$) distribution with (a)~log-normal
and (b)~power-law forms.
The parameters for the best-fit log-normal distribution ($\mu$ and
$\sigma$) and power-law distribution ($\alpha$) are
obtained from maximum likelihood estimation. The value of $G_T^{min}$
(in currency units: INR for India and USD for USA \& Japan) indicates
the minimum total gross beyond which log-normal or power-law fitting 
is applied to the tail of the empirical distribution. The value of 
$G_T^{min}$ is chosen 
such that the empirical and best-fit distributions are as
similar as possible in terms of minimum KS statistics.\\}
\label{table1}
\begin{tabular}{|lc|cccc|ccc|}
\hline
Country & Year(s) & \multicolumn{4}{c|}{Log-normal distribution} &
\multicolumn{3}{c|}{Power-law distribution}\\
\cline{3-9}
&  & $\mu$ & $\sigma$ & $G_T^{min}$ & $p$-value & $\alpha$ &
$G_T^{min}$ & $p$-value\\
\hline
& 1999 & 16.446 & 1.427 & & 0.786 & 3.953 & $12.69 \times 10^7$ &
0.956 \\
& 2000 & 16.574 & 1.484 & & 0.452 & 2.554 & $4.88 \times 10^7$ &
0.001  \\
& 2001 & 16.615 & 1.486 & & 0.517 & 2.553 & $5.02 \times 10^7$ &
0.192 \\
& 2002 & 16.569 & 1.468 & & 0.779 & 3.465 & $11.67 \times 10^7$ &
0.784 \\
{USA} & 2003 & 16.708 & 1.480 & ~$8 \times 10^6$~ & 0.248 &  3.519 &
$~9.56 \times 10^7~$ & 0.336 \\
& 2004 & 16.659 & 1.511 & & 0.542 & 2.751 & $5.71 \times 10^7$ &
0.666 \\
& 2005 & 16.702 & 1.434 & & 0.333 & 2.628 & $4.56 \times 10^7$ &
0.489 \\
& 2006 & 16.577 & 1.469 & & 0.115 & 2.848 & $5.23 \times 10^7$ &
0.285 \\
& 2007 & 16.567 & 1.477 & & 0.670 & 2.093 & $2.85 \times 10^7$ &
0.012 \\
& 2008 & 16.657 & 1.488 & & 0.242 & 3.833 & $12.75 \times 10^7$ &
0.800 \\
\hline
India & 1999-2008 & 18.264 & 0.922 & $1 \times 10^7$ & 0.395 &
2.296 & $7.75 \times
10^7$ & 0.001 \\
\hline
Japan &  2002-2008 & 15.725 & 1.097 & $5 \times 10^6$ & 0.074 &
2.232 & $9.56 \times
10^6$ & 0.000 \\
\hline
\end{tabular}
\end{table*}

To further establish that the log-normal curve indeed describes the
tail of the income distribution better than a power-law, we have also
tested for the significance of the fits using Kolmogorov-Smirnov (KS) statistics~\cite{Clauset09} 
(Table~\ref{table1}). After calculating the KS statistic for the
empirical distribution
and the best-fit curve  
obtained by maximum likelihood estimation (for both log-normal and power-law distributions), 
we have generated ensemble of random samples having same size as the empirical data from the best-fit
distributions and the KS-statistic is calculated for each such sample.
The $p$-value is obtained by measuring
the fraction of samples whose KS-statistic is greater than that obtained from the empirical data and a
higher $p$-value indicates greater confidence in the fit with the theoretical distribution.
We note from Table~\ref{table1} that the tails of the distributions of total gross earned by
movies released in USA for each year during 1999-2008 is well-described by
a log-normal distribution as indicated by the relatively high $p$-values,
as are the tails of the aggregated distributions for India and Japan (the tails correspond
to movies earning in excess of 8 million USD for USA, 10 million INR for India and 5 million USD for
Japan). By contrast, we have to reject the hypothesis that the Indian and Japanese 
distributions can be described by a power-law tail as the corresponding $p \leq 10^{-3}$.
For USA, the annual distributions for most years between 1999-2008 appear to have high $p$-value 
when a power-law is fit to the tail (corresponding approximately to movies earning in excess of 50 million USD).
However, 
a power-law fit to the tail of the aggregate distribution of the entire US data (comprising movies that
earned more than 1.10 billion USD) with the maximum likelihood estimated
exponent $\alpha \simeq 3.26$ gives a $p$-value of only about 0.024.
Thus, as the power-law form agrees with data from only one of the
three countries considered, and moreover, fits a smaller region
of the tail of the empirical distributions compared to the log-normal
curve, we consider the latter to be a more suitable choice for
describing the income distribution of movies than the former.

\begin{figure}
\begin{center}
  \includegraphics[width=0.99\linewidth]{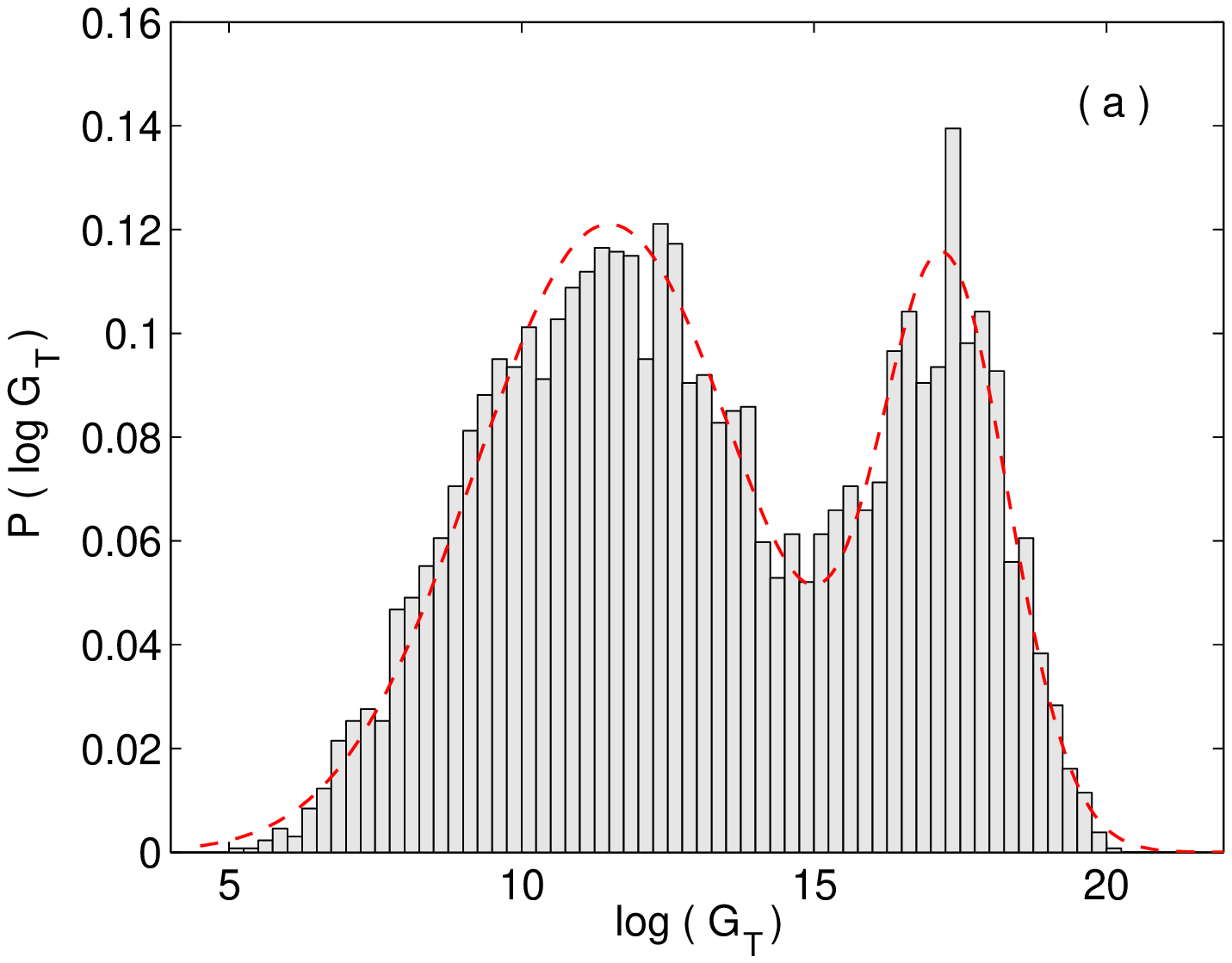}
  \includegraphics[width=0.99\linewidth]{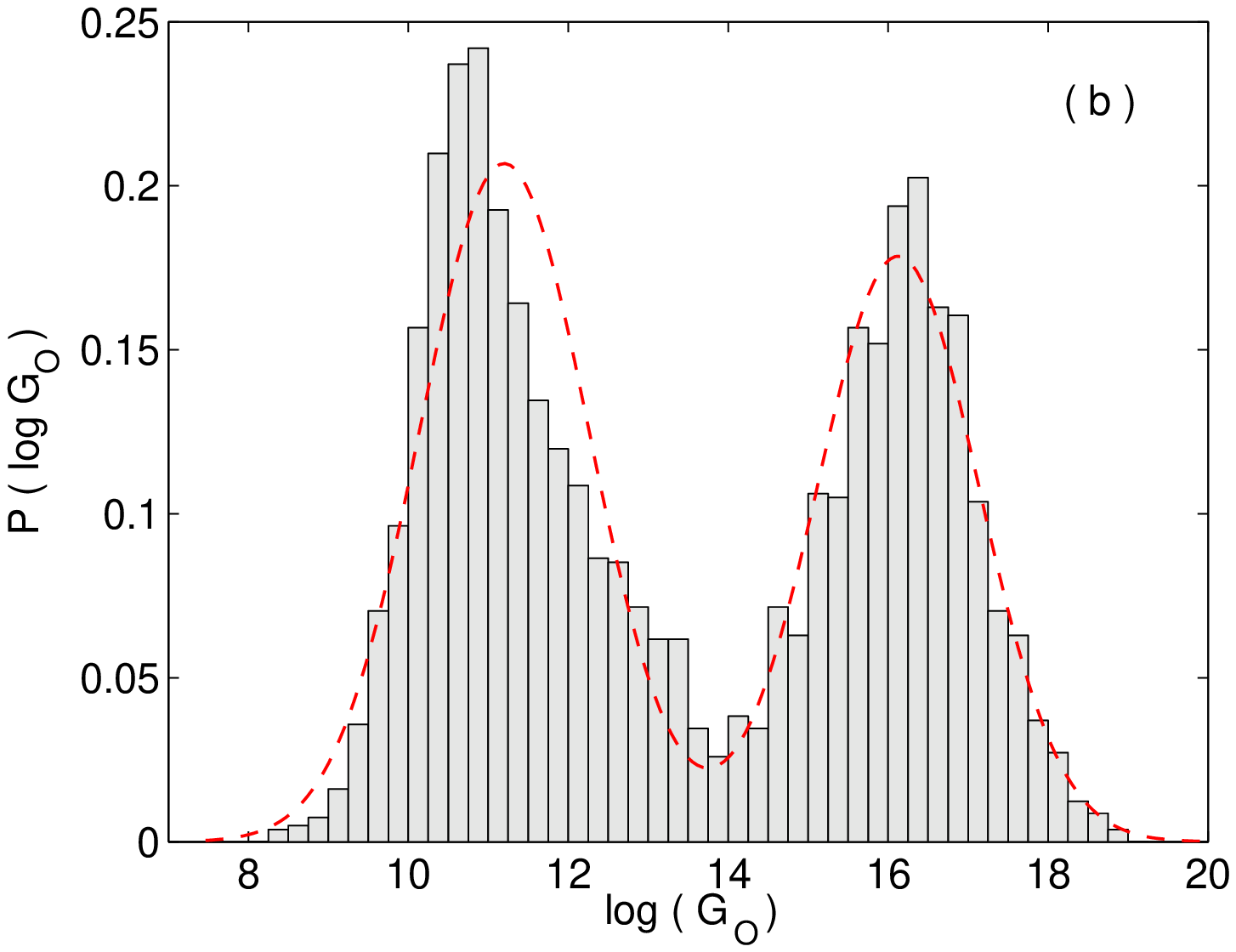}
\end{center}
\caption{
The logarithmically binned probability distributions for (a) the total
gross, $G_T$,
and (b) the opening gross, $G_O$, for movies released across theaters in USA 
during the period 1999-2008. Both distributions exhibit bimodal nature and have
been fit using a superposition of two log-normal distributions ${\cal
P}(x,\mu_1,\sigma_1)$
and ${\cal P}(x,\mu_2,\sigma_2)$ having relative weights $p$ and $(1-p)$, respectively.
For the total gross, the best fit is obtained using $p=0.699$,
$\mu_1=11.499$, 
$\mu_2=17.239$, $\sigma_1=2.305$ and $\sigma_2=1.089$, while for the opening
opening gross, the best fit parameters are $p=0.549$, $\mu_1=11.199$, 
$\mu_2=16.126$, $\sigma_1=1.058$ and $\sigma_2=1.008$.
Note that the data-set used for generating the distribution in (a) is
larger than that used for (b), as there are some movies
with low total gross whose opening week income is not available.
%
}
\label{fig:tgross_bimodal}
\end{figure}

Instead of focusing only at the tail (which corresponds to the top
grossing movies), if we now look at the entire income distribution, we
notice another important property: a bimodal nature.  There are two
clearly delineated peaks which corresponds to a large number of movies
either having a very low income or a very high income, with relatively
few movies that perform moderately at box-office
(Fig.~\ref{fig:tgross_bimodal}). The existence of this bimodality can
often mask the nature of the distribution, especially when one is
working with a small data-set. For example, De Vany and Walls, based
on their analysis of the gross for only about 300 movies, had stated
that log-normality could be rejected for their sample~\cite{Vany96}.
However, this assumed that the underlying distribution can be fit by a
single unimodal form - an assumption that was quite clearly incorrect
as evident from the histogram of their data.  Our more detailed and
comprehensive analysis with a much larger data-set shows that the
distribution of the total gross is in fact a superposition of two
different log-normal distributions:
\begin{equation} 
P (x) = p {\cal P}(x,\mu_1,\sigma_1) + (1-p) {\cal P}(x,\mu_2,\sigma_2),
\label{eq:bimodal}
\end{equation}
where ${\cal P} ( )$ represents the log-normal distribution while $p$ and $(1-p)$
are the relative weights of the two component distributions about the lower
and higher income peaks, respectively. We observe in
Fig.~\ref{fig:tgross_bimodal}
that the best-fit for the total income data occurs when $p \simeq 0.7$. 

Turning our focus now to the opening gross, we notice
that its distribution also has a bimodal form which can be similarly 
represented as a 
superposition of two log-normal distributions with the relative weights
of the two components being $p=0.55$ and $(1-p)=0.45$. The more
equal contributions of the distributions about the lower and 
higher income peaks in the opening week, as compared to that for 
the total or aggregated income over the entire lifetime seen earlier, 
is possibly because many movies which open with high viewership rapidly
decay in terms of popularity and exit from the cinemas within a few weeks. Thus,
their short lifetime translates into poor performance in terms of the
overall income and they do not contribute to the higher income peak
for the bimodal total gross distribution. However,
the total and opening gross distributions are qualitatively similar,
suggesting that the nature of the popularity distribution of movies is
decided at the opening week itself~\footnote{Note that this is a
statement about the {\em distribution} that pertains to the entire ensemble
rather than about an individual movie. The result does not imply that the gross
income of a particular movie on the opening week completely 
determines its total income.}. In this context, it is interesting to
note the recent finding that the long-term popularity of online content 
in web portals such as {\em YouTube}
can be predicted to a certain extent by observing the access
statistics of an item (e.g., a video) in the initial period after it has been
posted by an user~\cite{Szabo10}.

\subsection{Distribution of Gross per Theater}
\begin{figure}
\begin{center}
  \includegraphics[width=0.99\linewidth]{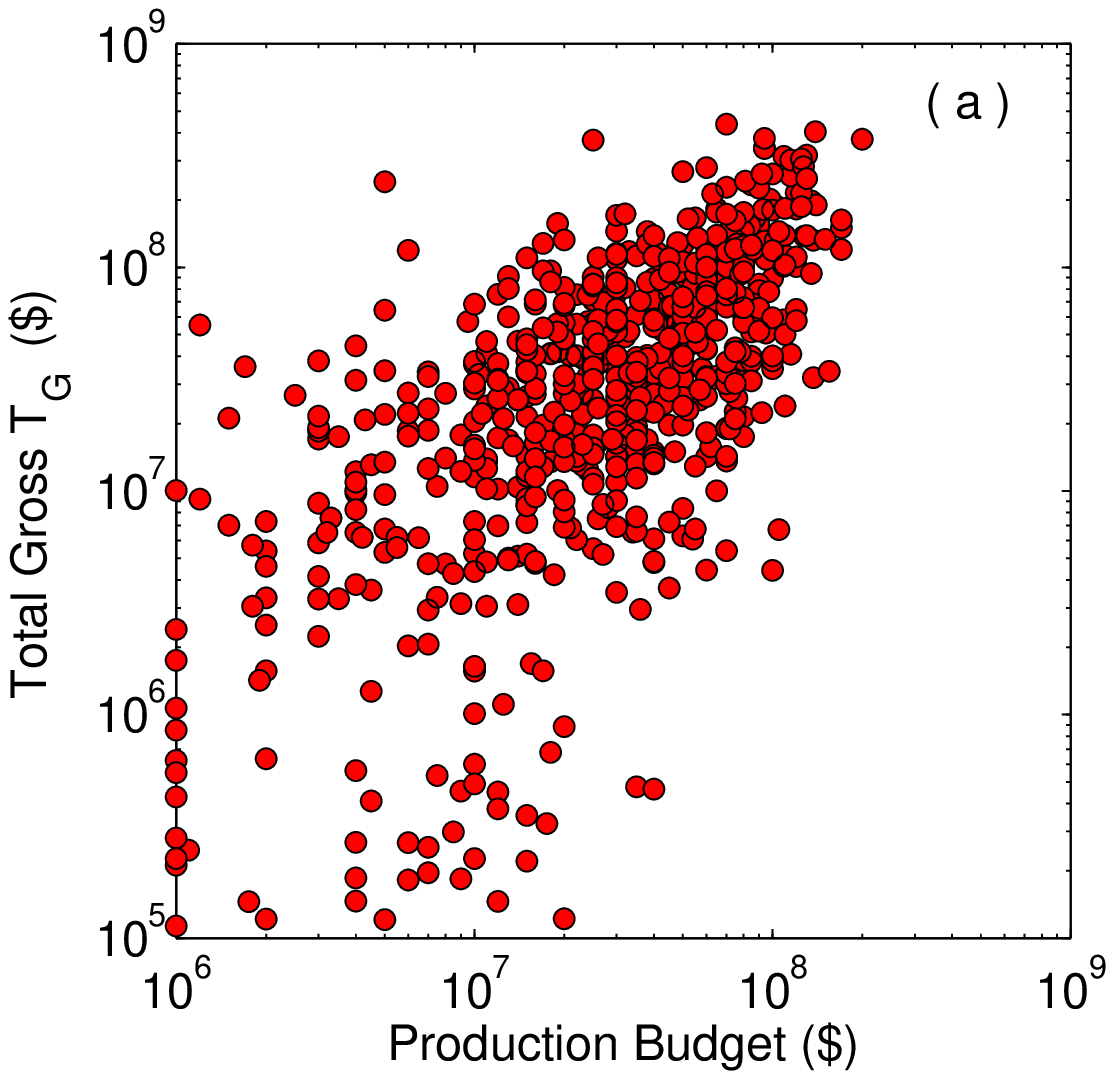}
  \includegraphics[width=0.99\linewidth]{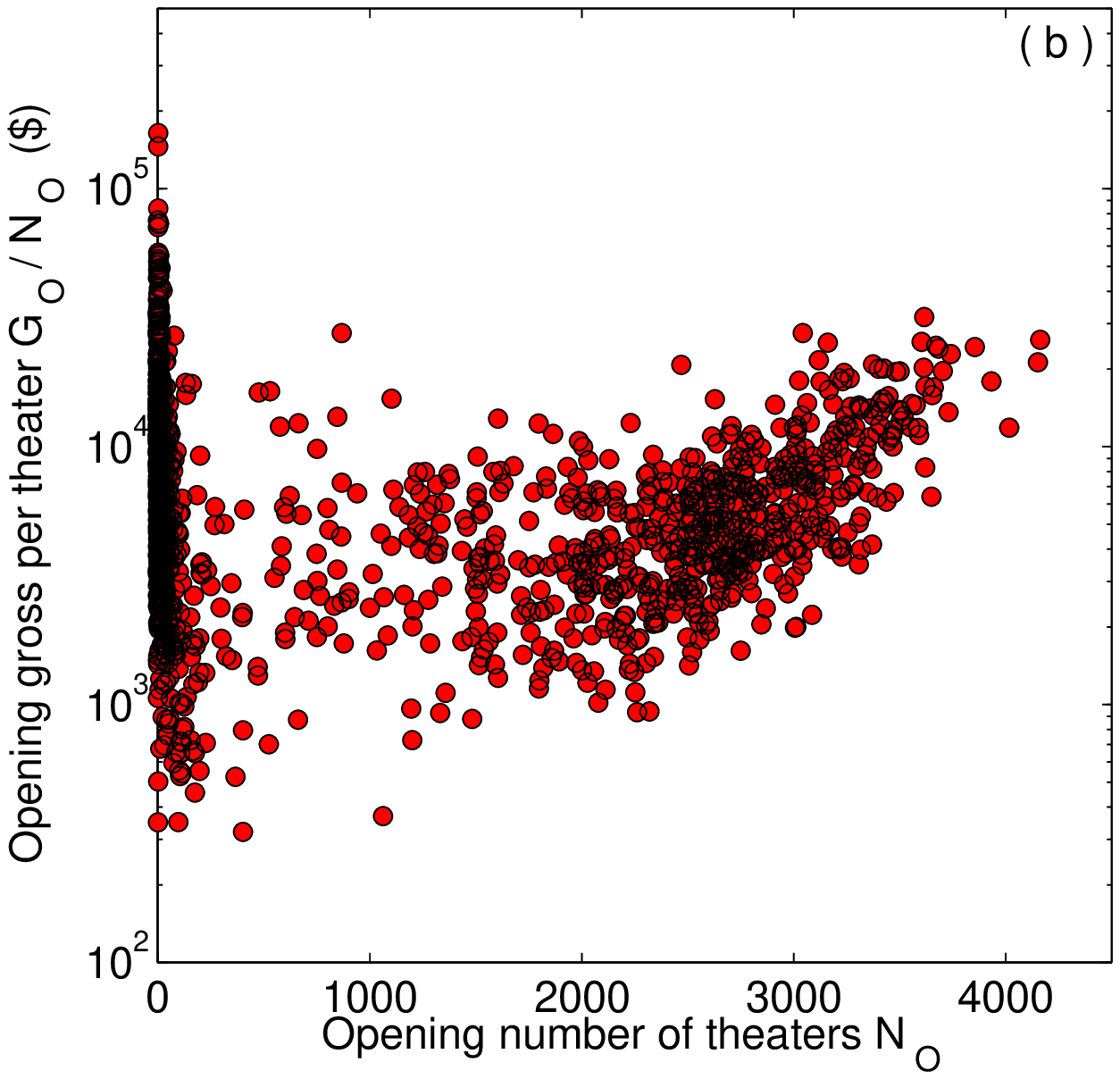}
\end{center}
\caption{
(a) Total gross ($G_T$) of a movie shown as a function of its
production budget. The correlation coefficient is $r = 0.633$ and 
the measure of significance $p < 0.001$. (b) The opening gross per theater
($g_O = G_O/N_O$)
as a function of the number of theaters in which a movie is released on the opening
week.
The correlation coefficient $r = -0.151$, with the corresponding 
measure of significance $p < 0.001$.
}
\label{fig:budget}
\end{figure}

We now focus on understanding what is responsible for the bimodal log-normal
distribution of the gross income for movies. It is, of course, possible that
this is directly related to the intrinsic quality of a movie or some other
attribute that is intimately connected to a specific movie (such as, how
intensely a film is promoted in the media prior to its release). Lacking any
other objective measure of the quality of a movie, we have used its
production budget as an indirect indicator. This is because movies with
higher budget would tend to have more well-known actors, better visual effects
and, in general, higher production standards. 
As mentioned earlier, we have considered movies for which publicly available information
about the production budget is available. This may not be the exact total cost 
incurred in making the movie, but nevertheless, gives an overall idea
about the expenses involved. 
Fig.~\ref{fig:budget}~(a) shows a scatter plot of the total gross as a function
of the
production budget for movies released between 1999-2008
whose budget exceeded 1 million dollars. As is clear from the figure, although in
general, movies with higher production budget do tend to earn more, the
correlation is not very high (the correlation coefficient $r$ is only 0.63). 
Thus, production budget by itself is not enough to guarantee high
popularity.~\footnote{Information about the production
budget of movies having low income is often not available. 
We have thus not been 
able to investigate whether the distribution of production budgets
itself has a bimodal nature.}

Another possibility is that the immediate success of a movie after its
release
is dependent on how well the movie-going public has been made aware
of the film by pre-release advertising through various public media. 
Ideally, an objective measure for this could be the 
advertising budget of the movie. 
However, as this information is mostly unavailable, 
we have used as a surrogate the
data about the number of theaters that a movie is initially released at.
As opening a movie at each theater requires organizing publicity for 
it among the neighboring population, and, wider release also implies more
intense mass-media campaigns, we expect the advertising cost to roughly scale with 
the number of opening theaters. As is obvious
from Fig.~\ref{fig:budget}~(b), the correlation between the opening gross 
per theater and the total number of theaters that a movie opens in is essentially
non-existent (the correlation coefficient $r \simeq -0.15$),
suggesting that advertising may not be a decisive factor for the
the success of a movie at the box-office. In this context,
one may note that De Vany \& Walls have looked at the distribution of movie
earnings and profit as a function of a variety of variables, such as,
genre, ratings, presence of stars, etc. and have not found any of these to
be significant determinants of movie performance~\cite{Vany03}. 

Instead of focusing on factors inherent to specific movies that can be used
to explain the gross distribution, we will now see
whether the bimodal log-normal nature appears as a result of two independent
factors, one responsible for the log-normal form of the component distributions
and the other for the bimodal nature of the overall distribution.
First, turning to the log-normal form, we observe that it may
arise from the nature of the distribution of gross income of a movie
normalized by the number of theaters in which it is being shown.
The {\em income per theater} gives us a more detailed view of the
popularity of a movie, compared to its gross aggregated over all
theaters.
It allows us to distinguish between the performance of two movies that
draw similar number of viewers, even though one may be shown at a much
smaller number of theaters than the other. This implies
that the former is actually attracting relatively larger audiences
compared to the
other at each theater, and hence, is more popular locally. Thus, the less
popular movie is generating the same income 
simply on account of it being shown in many more theaters, even though fewer
people in each locality served by the cinemas may be going to see it.

\begin{figure}
\begin{center}
  \includegraphics[width=0.99\linewidth]{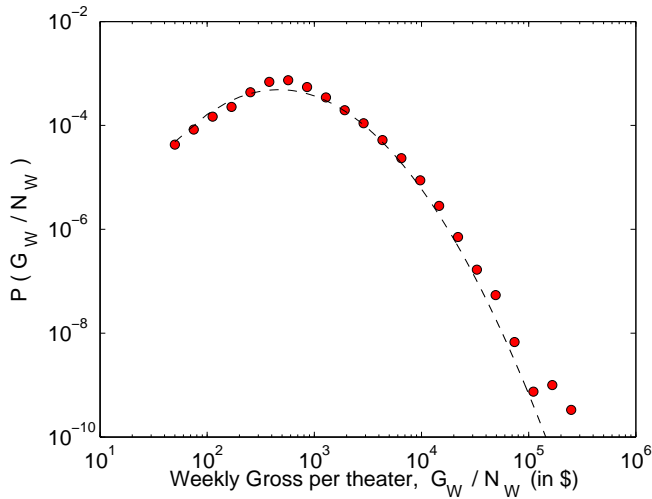}
\end{center}
\caption{The probability distribution of $g_W (= G_W/N_W)$, 
the gross income in a given
week per theater for a movie released in USA during the period
2000-2004 fit by a log-normal distribution (broken curve).
The parameters of the best-fit distribution are
$\mu=7.208 \pm 0.011$   and $\sigma=1.035 \pm 0.008$. The data is
averaged over all the weeks in the period mentioned.
}
\label{fig:gross_snapshot}
\end{figure}

\begin{figure}
\begin{center}
  \includegraphics[width=0.99\linewidth]{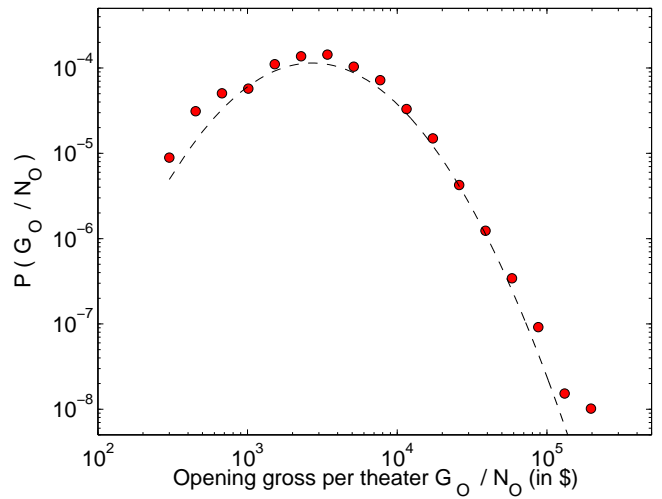}
\end{center}
\caption{The probability distribution of $g_O (=G_O/N_O)$, 
the opening week gross income per theater
of a movie released in theaters across USA during the period 2000-2004 fit by
a log-normal distribution (broken curve). The best-fit distribution parameters are
$\mu=8.672 \pm 0.044$   and $\sigma= 0.877\pm0.030$.
}
\label{fig:ogross_pertheater}
\end{figure}

Fig.~\ref{fig:gross_snapshot} shows that the distribution of the gross
income per theater over any given week, $g_W=G_W/N_W$,
has a log-normal nature. 
Here, $G_W$ represents the total gross income of a movie over a week
$W$ when it is being shown at $N_W$ theaters.
Note that this is quite different 
from the earlier distributions because we are now considering together movies 
that are at very different stages in their lifetime. On a particular week,
a few movies might have just opened, others are about to be withdrawn from exhibition, 
while still others are somewhere in the middle of their theatrical run.
On the other hand, Fig.~\ref{fig:ogross_pertheater} shows the distribution
of the income per theater of a movie on its opening week, i.e.,
$g_O = G_O/N_O$, where $N_O$ is the number of theaters in which the movie
is released. 
Calculating the KS statistics for the log-normal fitting of the opening gross per theater
gives a $p$-value of 0.678, indicating the fit to be statistically significant.
Thus, both the distribution of $g_W$ and that of $g_O$ show a log-normal form despite 
the fact that they are quite different quantities, thereby underlining the robustness
of the nature of the distribution.  

The appearance of the log-normal distribution may not be surprising 
in itself,
as it is expected to occur in any linear multiplicative
stochastic process~\cite{Mitzenmacher03,Ciuchi93}.
The decision to see a movie (or not) can be considered to be
the result of a sequence of independent choices, each of which have certain
probabilities. Thus, the final probability that an individual will go to the
theater to watch a movie is a product of each of these constituent probabilities,
which implies that it will follow a log-normal distribution.
It is worth noting here that the log-normal distribution also appears in 
other areas where popularity of different entities arise as a result
of collective decisions, e.g., in the context of proportional
elections~\cite{Fortunato07}, citations of scientific
papers~\cite{Redner04,Radicchi08} and visibility of news stories posted 
by users on an online web-site~\cite{Wu07}.

\subsection{Distribution of the number of theaters a movie is released}
\begin{figure}
\begin{center}
  \includegraphics[width=0.99\linewidth]{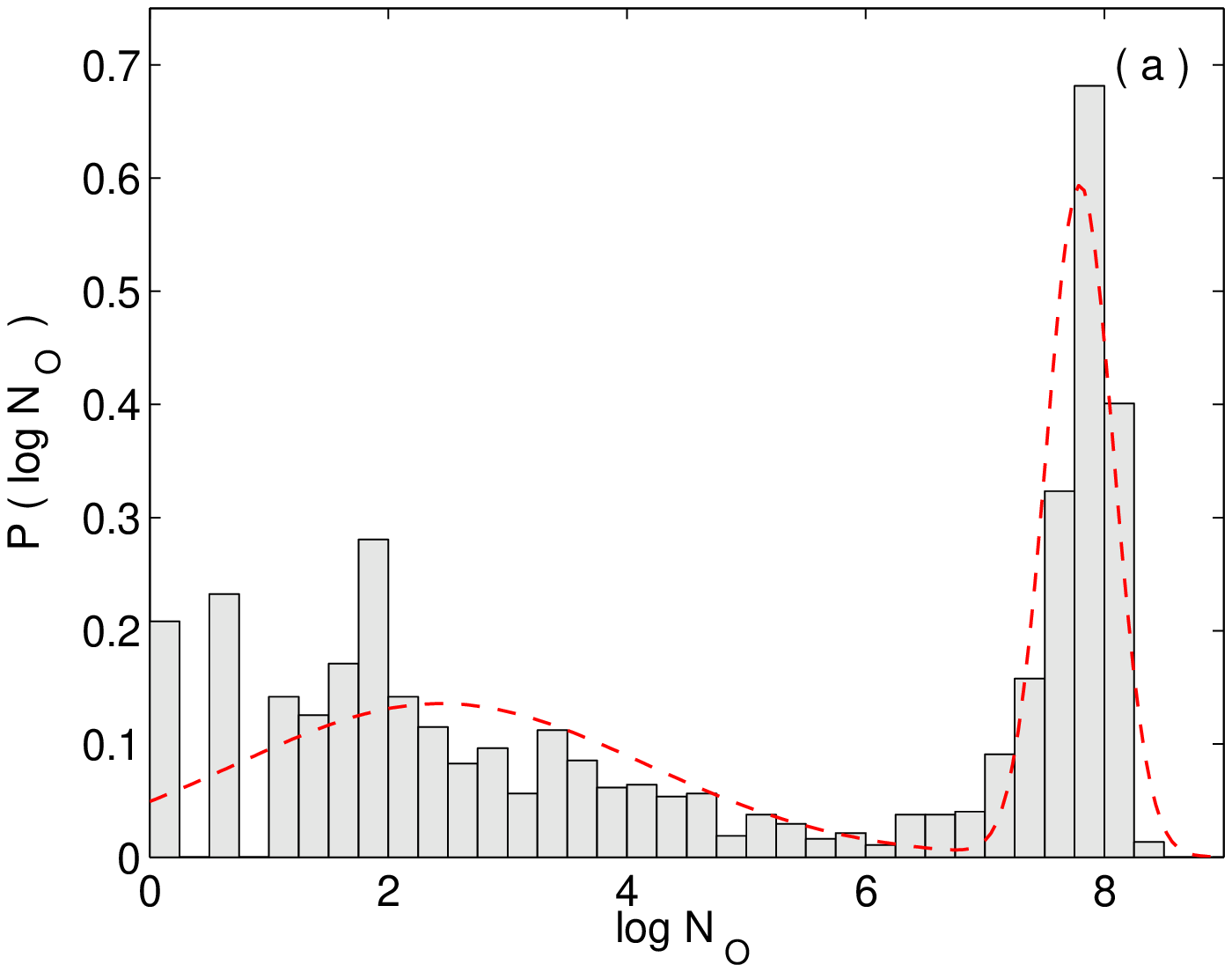}
  \includegraphics[width=0.99\linewidth]{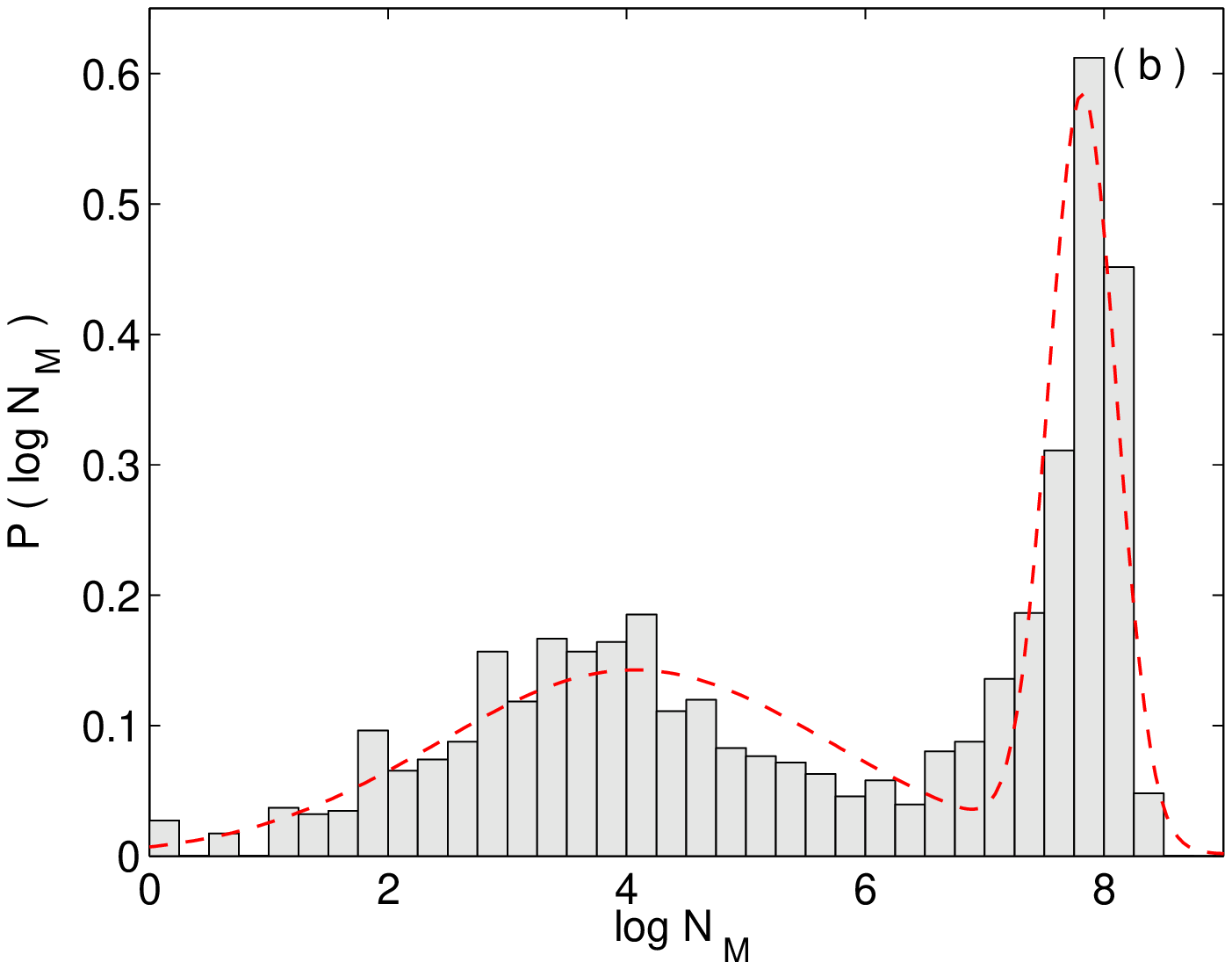}
\end{center}
\caption{The probability distribution of the logarithm of
(a) the number of theaters in which a movie is shown on its opening
week, $N_O$, and (b) the largest number of theaters, $N_M$, 
that a movie is simultaneously shown on a given week in its entire lifetime. 
The data is for
movies released across USA during 2000-2004. A
fit using two Gaussian distributions is shown for comparison.
The parameters of the best-fit distribution for the number of theaters in the opening week are
$p=0.583$, $\mu_1=2.442$, $\mu_2=7.796$, $\sigma_1=1.714$ and
$\sigma_2=0.281$,
while that for the largest number of theaters in a week over its lifetime are 
$p=0.590$, $\mu_1=4.066$, $\mu_2=7.821$, $\sigma_1=1.651$ and
$\sigma_2=0.285$.
}
\label{fig:otheaters_bimodal}
\end{figure}

While the log-normal nature of the popularity distribution for movies
(as measured by their gross income per theater)
can be explained as the consequence of an underlying process where
sequential stochastic choices are made by each individual to decide
whether to watch the movie, its overall bimodal character is yet to be
explained.
Fig.~\ref{fig:otheaters_bimodal} suggests a possible answer: the two peaks of
the distribution of income may be reflecting the bimodality of the
distribution of $N_O$, i.e., the number of theaters in which a motion 
picture is released, or of $N_M$, 
the largest number of cinemas that it plays simultaneously
at any point during its entire lifetime. Thus, most movies are shown either at 
a handful of theaters, typically a hundred or less (these are usually the independent or foreign movies) or
at a very large number of cinema halls, numbering a few thousand (as
is the case with the products of major Hollywood studios).
Unsurprisingly, this also decides the overall popularity of the movies
to an extent, as the potential audience of a film running in less than
a hundred theaters is always going to be much smaller than what we
expect for blockbuster films. In most cases, the former will be much smaller
than the critical size required to generate a positive word-of-mouth
effect spreading through mutual acquaintances which will gradually
cause more and more people to become interested in seeing the film.
There are occasional instances where such a movie does manage to
make the transition successfully, when a major distribution house,
noticing an opportunity, steps in to market the film nationwide to a
much larger audience and a ``sleeper hit'' is created.  
An example is
the movie {\em My Big Fat Greek Wedding}, that opened in only 108
theaters in 2002 but went on to become the fifth highest grossing
movie for that year, running for 47 weeks and at its peak was shown in
more than 2000 theaters simultaneously.

Bimodality has also been observed in other popularity-related contexts, such as, 
in the electoral dynamics of US Congressional elections, where
over time the margin between the victorious and defeated candidates have been
growing larger~\cite{Mayhew74}. For instance, the proportion of votes won by the
Democratic Party candidate in the federal elections has changed from around
50$\%$ to one of two possibilities: either around 35-40$\%$ (in which case the
candidate lost) or around 60-65$\%$ (when the candidate won). We have earlier
proposed a theoretical framework for explaining how bimodality can arise in
such collective decisions arising from individual binary choice
behavior~\cite{Pan06,Sinha06a,Sinha06b}. 
Individual agents took ``yes'' or ``no'' decisions on issues based on information
about the decisions taken by their neighbors and were also influenced by their own
previous decisions (adaptation), as well as, how accurately their neighborhood had
reflected the majority choice of the overall society in the past (learning).
Introducing these effects in the evolution of preferences for the agents led
to the emergence of two-phase behavior marked by transition from a unimodal
behavior to a bimodal distribution of the fraction of agents favoring a particular choice, as
the parameter controlling the learning or global feedback is increased. 
In the context of the movie income data, we can identify this choice dynamics as
a model for the decision process by which theater owners and movie distributors agree to release
a particular movie in a specific theater. The procedure is likely to be
significantly influenced by
the previous experience of the theater and the distributor, as both learn from
previous successes and failures of movies released/exhibited by them in the past,
in accordance with the assumptions of the model. Once released in a theater,
its success will be decided by the linear multiplicative stochastic process outlined earlier
and will follow a log-normal distribution.
Therefore, the total or opening gross distribution for movies may be
considered to be a combination of the log-normal distribution of
income per theater and the bimodal distribution of the number of
theaters in which a movie is shown.

\subsection{Time-evolution of movie popularity}
\begin{figure}
\begin{center}
  \includegraphics[width=0.99\linewidth]{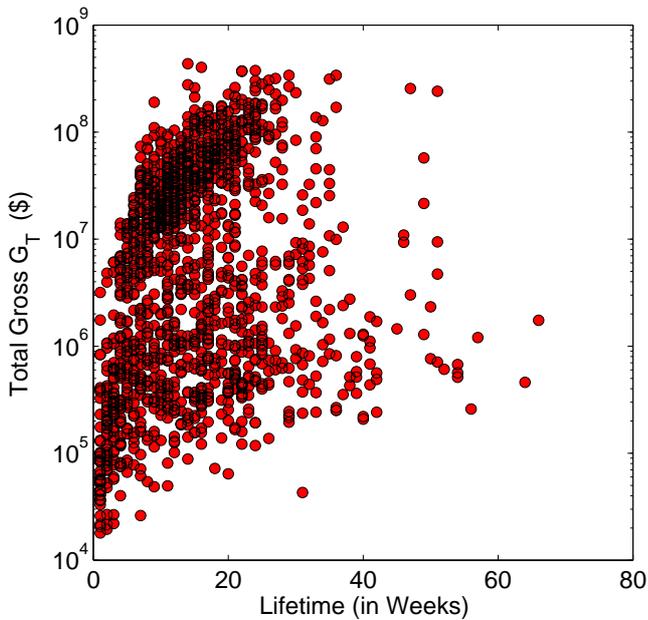}
\end{center}
\caption{
The total gross ($G_T$) earned by a movie as a function of its lifetime,
i.e., the duration of its run at theaters.
The correlation coefficient is $r = 0.224$ with the corresponding
measure of significance $p<0.001$.
}
\label{fig:theater_vs_tgross}
\end{figure}

The distribution of gross income analyzed so far represents a
temporally aggregated view of the popularity of movies. As already
mentioned, information about how the number of viewers change over
time can reveal other aspects of the popularity dynamics, and in
particular, can distinguish between movies in terms of how they have
achieved their success at the box-office, viz., blockbusters and
sleepers. This can be seen by looking at a plot of the total income of
movies against their income
in the opening week, with blockbusters having high values for both
while sleepers would be characterized by a low opening but high total
gross. One can also observe distinct classes of movies from the
scatter-plot shown in
Fig.~\ref{fig:theater_vs_tgross} which indicates the correlation
between the total gross income of a movie, representing its overall
popularity, and its lifetime at the theaters, which is a measure for
how long it manages to attract a sufficient number of viewers. We
immediately notice that although many movies with high total gross
also tend to have long lifetimes, there are also several movies which
lie outside this general trend.  Films which tend to have a very long
run at the theaters despite not having as high a total income as major
studio blockbusters may belong to a special class, e.g., effects
movies which are made to be shown only at theaters having giant
screens~\cite{Sinha04}.

\begin{figure}
\begin{center}
  \includegraphics[width=0.99\linewidth]{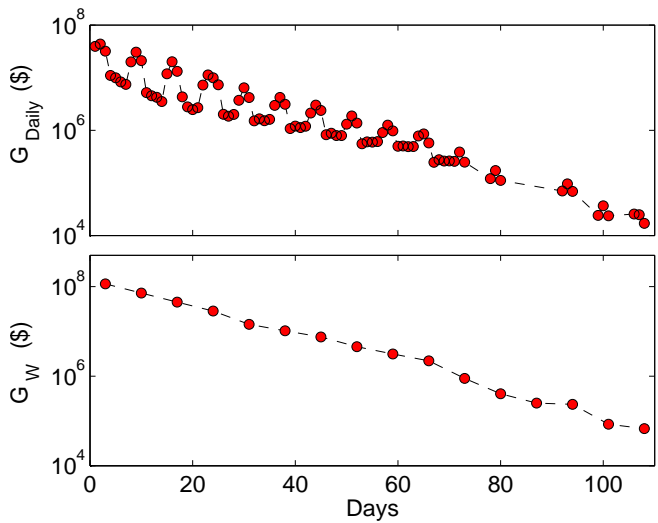}
\end{center}
\caption{The time-evolution of (top) the daily gross income,
$G_{Daily}$ (in dollars), and (bottom) the
weekly gross income $G_W$ (in dollars) for the film {\em Spiderman} (released in 2002) 
aggregated over all theaters across USA. The gross income is seen to
decay with time approximately exponentially, i.e., $\sim \exp(-\gamma
t)$,
where $\gamma \simeq 0.072$/day, giving a characteristic decay time of 14 days
(i.e., 2 weeks).
}
\label{fig:daily_gross}
\end{figure}

\begin{figure}
\begin{center}
  \includegraphics[width=0.99\linewidth]{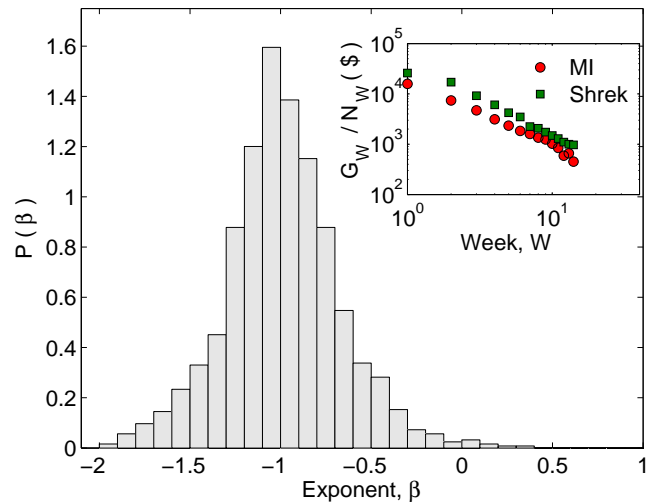}
\end{center}
\caption{The distribution of exponents $\beta$ describing the power-law decay
of the weekly gross income per theater, $g_W$ ($= G_W/N_W$) for all movies
released in USA during 2000-2004 which ran in theaters for more than 5 weeks. 
The exponents were obtained by least-square linear fitting 
of the weekly gross income per theater of a movie as a function of time (measured
in weeks, $W$) on a doubly-logarithmic scale. The inset shows the decay for two
specific movies, {\em Mission Impossible (1996)} and {\em Shrek} (2001).
}
\label{fig:power_law}
\end{figure}

To go beyond the simple blockbuster-sleeper distinction and have a
more detailed view of the time-evolution of movie popularity, one has
to consider the trend followed by the daily or weekly income of a
movie over time. Fig.~\ref{fig:daily_gross}~(top) shows that the daily
gross income of a classic blockbuster movie, {\em Spiderman} (released
in 2002) decays with time after release, having regularly spaced peaks
corresponding to large audiences on weekends. To remove the intra-week
fluctuations and observe the overall trend, we focus on the time
series of weekly gross, $G_W$ (Fig.~\ref{fig:daily_gross},~bottom). This
shows an exponential decay with a characteristic rate $\gamma$, 
a feature seen not only for almost all
other blockbusters, but for bombs as well. The only difference between
blockbusters and bombs is in their initial, or opening, gross.
However, sleepers may behave differently, showing an initial increase
in their weekly gross and reaching the peak in the gross income
several weeks after release. For example, in the case of the 
movie {\em My Big Fat Greek Wedding}, the peak occurred 20 weeks after 
its initial opening. It is then followed by exponential decay of the
weekly gross until the movie is withdrawn from circulation. Note that
the exponential decay rate ($\gamma$) is different for different 
movies~\footnote{In general, it is also possible for popularity dynamics
to exhibit bursts or avalanches, as has indeed been observed in the context of
the popularity of online documents (measured both in terms of the number of
hyperlinks pointing to a webpage and the number of visits made to the
page)~\cite{Ratkiewicz10}. 
However, we have not observed any significant burst-like
pattern in the dynamics of movie popularity.}.

Instead of looking
at the income aggregated over all theaters, if we consider the
weekly gross income per theater, a surprising universality is observed. 
As previously mentioned, the income per theater
gives us additional information about the movie's popularity
because a movie that is being shown in a large number of theaters may
have
a bigger income simply on account of higher accessibility for the
potential audience. Unlike the overall gross that decays exponentially
with time, the gross per theater of a movie shows a power-law decay in time
measured in terms of the number of weeks from its release, $W$:
$g_W \sim W^{-\beta}$, with exponent 
$\beta \simeq -1$~\cite{Sinha05} (see the inset of Fig.~\ref{fig:power_law}).
This has a striking similarity with the
time-evolution of popularity for scientific papers in terms of citations.
It has been reported that the citation probability to a paper published $t$
years ago, decays approximately as $1/t$~\cite{Redner04}.
Note that, Price~\cite{Price76} had
also noted a similar behavior for the decay of citations to papers listed
in the Science Citation Index.  In a very different context, namely, the
decay over time in the popularity of a website (as measured by the rate 
of download of papers from the site) and that of individual
web-pages in an online news and entertainment portal (as measured by the 
number of visits to the page), power laws have also been reported
but with different exponent~\cite{Sornette00,Dezso06}.
More recently, relaxation dynamics of popularity with a power law
decay have
been observed for other products, such as,
book sales from {\em Amazon.com}~\cite{Sornette04} and the daily views of videos
posted on {\em YouTube}~\cite{Sornette08}, where the exponents
appear to cluster around multiple distinct classes.

Fig.~\ref{fig:power_law} shows the distribution of the power-law scaling exponents
for all movies which ran for more than 5 weeks in theaters in USA during the
period 2000-2004. 
Most of the movies have negative exponents, the mean and median being
$-0.990$ and $-1.002$, 
respectively, indicating a monotonic decay in the income per theater as a reciprocal
function of the time elapsed from the initial release date. 
Note that only 9 movies over the entire period analyzed by us had
positive exponents and
these had been shown at a very small number of theaters (around $5$ or less). 
For these movies the income per theater showed a slight increase towards the end of their
lifetime before they were completely pulled out from theaters.  
On the whole, therefore, the local popularity of a movie at a certain point
in time appears to be inversely proportional to the duration that has
elapsed from its initial release.

\section{Discussion}
\begin{figure}
\begin{center}
  \includegraphics[width=0.99\linewidth]{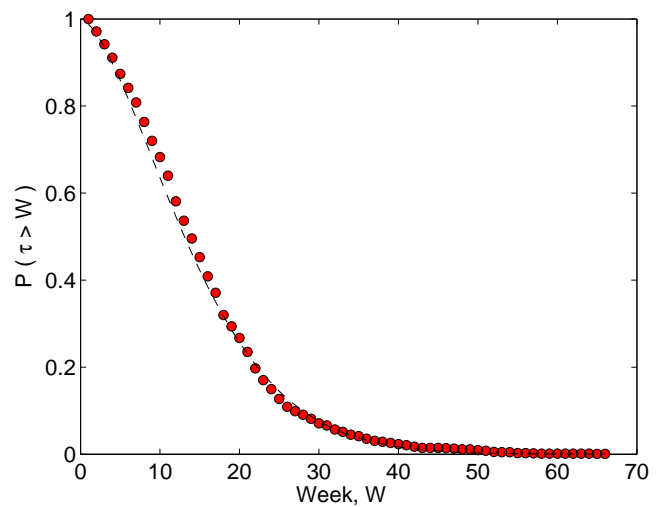}
\end{center}
\caption{The cumulative persistence probability of a movie to remain
in a theater for a period exceeding $W$ weeks shown as a function of
time ($W$, in weeks) for 
movies released in USA during 2000-2004. The broken line shows a fit with
the stretched exponential distribution (see text).
}
\label{fig:persistence}
\end{figure}

In the preceding section we have summarised the principal results from
our analysis of the data on movie popularity as reflected in their gross
income at the box office. One can now ask whether the three main features
observed, viz., (i) log-normal distribution of gross per theater, (ii) 
bimodal distribution of the number of theaters a movie is shown at 
and (iii) the decay in gross per theater of a movie as an inverse function
of the duration from its initial release, are sufficient to explain all
other observable properties of movie popularity. To illustrate this
we now consider an important quantifier not considered earlier: the
{\em persistence time} of a movie, $\tau$, i.e., the duration upto which it
is shown at theaters. As seen from Fig.~\ref{fig:persistence}, only about
half of all the movies considered survive for beyond 14 weeks in theaters
and only about $10 \%$ persist beyond 25 weeks. The cumulative
distribution fits a {\em stretched exponential} form ($\sim {\rm
exp} [- (t/a)^b ]$), indicating that
the persistence time probability distribution can be described by the Weibull 
distribution~\cite{Weibull51}: 
\begin{equation} 
  P(t) = \frac{b}{a} \left(\frac{t}{a}\right)^{(b-1)}
  \exp\left[-\left(\frac{t}{a}\right)^b\right],
\label{eq:weibull}
\end{equation}
where $a, b >0 $ are the shape and scale parameters of the distribution
respectively. The best fit to the data shown in Fig.~\ref{fig:persistence} 
is achieved for $a=16.485 \pm 0.547$ and $b=1.581 \pm 0.060$.
The Weibull distribution is well-known in the study of failure
processes and is often used to describe extreme events or large
deviations. In particular, it has been applied to describe the failure rate
of components, with the parameter $b > 1$ indicating that the rate increases
with time because of an aging process.

We can derive this empirical property of movie popularity from our earlier 
stated observations, with only an added assumption: that a movie is withdrawn
from circulation when its gross income per theater falls below a critical
value, $g_{c}$. An interpretation of this number is that it is related
to the minimum
number of tickets a theater has to sell per week in order to make the
exhibition of the movie an economically viable proposition. Once the popularity
of the movie has gone below this level, the theater is presumably no longer making
a profit by showing this film and is better off showing a
different movie. Therefore, the probability that the persistence time of 
a movie is $\tau$, is essentially given by the probability that the gross income
per theater at time $\tau$ falls below $g_c$.
To evaluate this probability, we use the observation that the gross
per theater of a movie $i$ at any time $t$, $g^i_t$, has decayed 
by a factor $1/t$ (on average) from its initial value, $g^i_O$, i.e.,
the opening gross per theater. As the decay of gross income is not a
completely deterministic process, we can express $g^i_t = (g^i_O / t)
\eta$, where $\eta$ has a log-normal distribution with parameters
$\mu$
($=0$) and $\sigma$, which guarantees that the stochastic variation can never
result in a negative value for the gross.
If time is expressed in units of weeks ($j=1,2,\ldots$), 
the cumulative probability distribution of the persistence time
$\tau_i$ is given by
\begin{equation}
P (\tau_i > T |g^i_O) = \prod^T_{j=1} P (g^i_j > g_c |g^i_O),
\label{eq:pers1}
\end{equation}
which is a product of the probabilities that the gross income per
theater earned by the movie in the $T$ successive weeks beginning from
its initial release are all greater than the critical value $g_c$.
We note that the right hand side expression of Eq.~\ref{eq:pers1} is equal to
$\prod^T_{j=1} P ( \eta > g_c~j/g^i_O |g^i_O)$, which is the product
of cumulative distribution functions for the
log-normal random variable $\eta$. Therefore, the cumulative distribution
of the persistence time can be written as:
\begin{equation}
P (\tau_i > T) = \int_{0}^{\infty} P(g^i_O) \prod^T_{j=1} \frac{1}{2}
\left(1 - {\rm erf} \left[\frac{{\rm
ln}(g_c j/g^i_O)}{\sqrt{2 \sigma^2}}\right] \right) dg^i_O,
\end{equation}
where $P(g^i_O)$ is the log-normal distribution of the opening gross
per theater and ${\rm erf} (x)$ is the error function.
Numerical solution of the above expression shows that it reproduces 
a stretched exponential curve, having reasonable agreement with
the empirical data.

The fact that at least certain aspects of popularity dynamics can be
described by the mathematical framework for understanding failure
events is probably not a coincidence. 
A common observation across different areas in which popularity
dynamics is operational is that most entities are pushed out of the
market within a short time of their introduction. In many areas, this
high rate of early extinction is balanced by new entrants into the
market, so that at any given time, the number of competing entities is
maintained more or less constant. 
Therefore, the key question in
understanding popularity is why do most
products or ideas fail to survive the
brutal competition for being the most popular choice. We see
instances of the general rule that ``most things fail'' in almost all
types of economic phenomena~\cite{Ormerod05}. As Ormerod has pointed out, of
the 100 top business enterprises is 1912, 48 had ceased to exist as
independent entities by 2005, while only 28 companies were larger (in real
terms) than they were back in 1912. 
Similarly, out of the large number of computer software companies
which had been in operation in the 1980s, only a handful have survived
to the present. 
This is not just true for the
present age but also for historical times. In 1469, twelve publishing
houses had emerged in Venice in the (at that time) 
pioneering activity of printing
books, but by the end of three years, only three of them had survived.
Thus, a successful product is
marked by its ability to survive in its early stages to build a
consumer base. This may be a product of chance, as often there is
little to distinguish between competitors in terms of their intrinsic
qualities. However, once a product or idea has had a certain number of
adherents, it is able to survive by the process of positive feedback,
thereby generating more adherents. In the case of movies, the process
reaches a natural limit when the potential audience is exhausted as
everybody likely to view the movie has seen the film. In
other cases, e.g., for religious or political ideas, the
process can continue indefinitely as more and more converts are added
to the fold. Thus, a general view of popularity dynamics can be
expressed as follows: each competing product or idea, upon initial
introduction to the market, has a certain probability of failing to
attract a sufficient number of adherents. This means that at every successive
time-step, the entity has to survive the possibility of early demise.
A popular object, according to this view, is one which has repeatedly
managed, by a combination of chance or design, to avoid failure
(and subsequent exit from the marketplace) for far longer than its
competitors.

\section{Conclusions: The stylized facts of popularity}

In this paper, we have analyzed the empirical data for movie
popularity, measured in terms of its box-office income. We observe
that the complex process of popularity can be understood, at least for
movies, in terms of three robust features which (using the
terminology of economics) we can term as {\em stylized facts of
popularity}: (i) log-normal distribution of the gross income per
theater, (ii) the bimodal distribution of the number of theaters in
which a movie is shown and (iii) power-law decay with time of the
gross income per theater. Some of these features have been seen in
other instances in which popularity dynamics plays a role, such as, in
citations of scientific papers or in political elections. This
suggests that it may be possible that the above three properties may
apply more generally to the processes by which a few entities emerge
to become a popular product or idea. A unifying framework may be
provided by the understanding of popular objects as those which have
repeatedly survived a sequential failure process.

\noindent
\begin{center}
{\bf Acknowledgments}
\end{center}

We would like to thank S Raghavendra, D Stauffer, J Kertesz, M Marsili
and D Sornette for helpful comments at various stages. 
We gratefully acknowledge discussions with S V Vikram regarding the 
theoretical calculations in section 5.
This work was
supported in part by the IMSc Complex Systems (XI Plan) Project.


\end{document}